\documentclass{ieeetj}
\usepackage{amsmath, amsfonts}
\usepackage{graphicx, color}
\usepackage[table]{xcolor}
\usepackage{hyperref}
\usepackage{multirow}
\usepackage{ragged2e}

\graphicspath{{./img/}}

\usepackage[backend=biber,hyperref,doi=false,isbn=false,style=ieee]{biblatex}
\defbibheading{bibliography}[]{}

\AtBeginDocument{
	\definecolor{tmlcncolor}{cmyk}{0.93,0.59,0.15,0.02}
	\definecolor{NavyBlue}{RGB}{0,86,125}
	\definecolor{tablecolor}{RGB}{51,133,255}
}

\def\OJlogo{\vspace{-4pt}$<$Society logo(s) and publication title will appear here.$>$}
\def\seclogo{\vspace{10pt}$<$Society logo(s) and publication title will appear here.$>$}
\def\authorrefmark#1{\ensuremath{^{\textbf{#1}}}}

\title{Bias-Switchable Row-Column Array Imaging using Fast Orthogonal Row-Column
       Electronic Scanning (FORCES) Compared with Conventional Row-Column Array Imaging}

\author{Randy Palamar\authorrefmark{1}, Graduate Student Member, IEEE,
        Mohammad Rahim Sobhani\authorrefmark{1}, Member, IEEE,
        Darren Dahunsi\authorrefmark{1}, Graduate Student Member, IEEE,
        Negar Majidi\authorrefmark{1}, Graduate Student Member, IEEE,
        Afshin Kashani Ilkhechi\authorrefmark{1}, Member, IEEE,
        Joy Wang\authorrefmark{1}, Graduate Student Member, IEEE,
        Jeremy Brown\authorrefmark{2}, Member, IEEE,
        and Roger Zemp\authorrefmark{1}, Member, IEEE}
\affil{Department of Electrical and Computer Engineering, University of Alberta, Edmonton, AB T6G 2R3, Canada}
\affil{School of Biomedical Engineering, Dalhousie University, Halifax, NS B3H 4R2, Canada}
\corresp{Corresponding authors: Randy Palamar (email: palamar@ualberta.ca),
         Roger Zemp (email: rzemp@ualberta.ca).}
\authornote{Funding for this work was provided by the National
Institutes of Health (1R21HL161626-01 and EITTSCA R21EYO33078),
Alberta Innovates (AICE 202102269, CASBE 212200391 and LevMax
232403439), MITACS (IT46044 and IT41795), CliniSonix Inc., NSERC
(2025-05274), the Alberta Cancer Foundation and the Mary Johnston
Family Melanoma Grant (ACF JFMRP 27587), the Government of Alberta
Cancer Research for Screening and Prevention Fund (CRSPPF 017061),
an Innovation Catalyst Grant to MRS, and INOVAIT (2023-6359). We
are grateful to the nanoFAB staff at the University of Alberta for
facilitating array fabrication.}

\receiveddate{XX Month, XXXX}
\reviseddate{XX Month, XXXX}
\accepteddate{XX Month, XXXX}
\publisheddate{XX Month, XXXX}
\currentdate{XX Month, XXXX}
\doiinfo{XXXX.2022.1234567}

\begin{document}

\begin{abstract}

Row-Column Arrays (RCAs) offer an attractive alternative to fully
wired 2D-arrays for 3D-ultrasound, due to their greatly simplified
wiring. However, conventional RCAs face challenges related to
their long elements. These include an inability to image beyond
the shadow of the aperture and an inability to focus in both
transmit and receive for desired scan planes. To address these
limitations, we recently developed bias-switchable RCAs, also
known as Top Orthogonal to Bottom Electrode (TOBE) arrays. These
arrays provide novel opportunities to read out from every element
of the array and achieve high-quality images. While TOBE arrays
and their associated imaging schemes have shown promise, they have
not yet been directly compared experimentally to conventional RCA
imaging techniques. This study aims to provide such a comparison,
demonstrating superior B-scan and volumetric images from two
electrostrictive relaxor TOBE arrays, using a method called Fast
Orthogonal Row-Column Electronic scanning (FORCES), compared to
conventional RCA imaging schemes, including Tilted Plane Wave
(TPW) compounding and Virtual Line Source (VLS) imaging. The study
quantifies resolution and Generalized Contrast to Noise Ratio
(gCNR) in phantoms, and also demonstrates volumetric acquisitions
in phantom and animal models.

\end{abstract}


\maketitle

\section{INTRODUCTION}

\begin{figure*}
	\centering
	\includegraphics[width=0.9\textwidth]{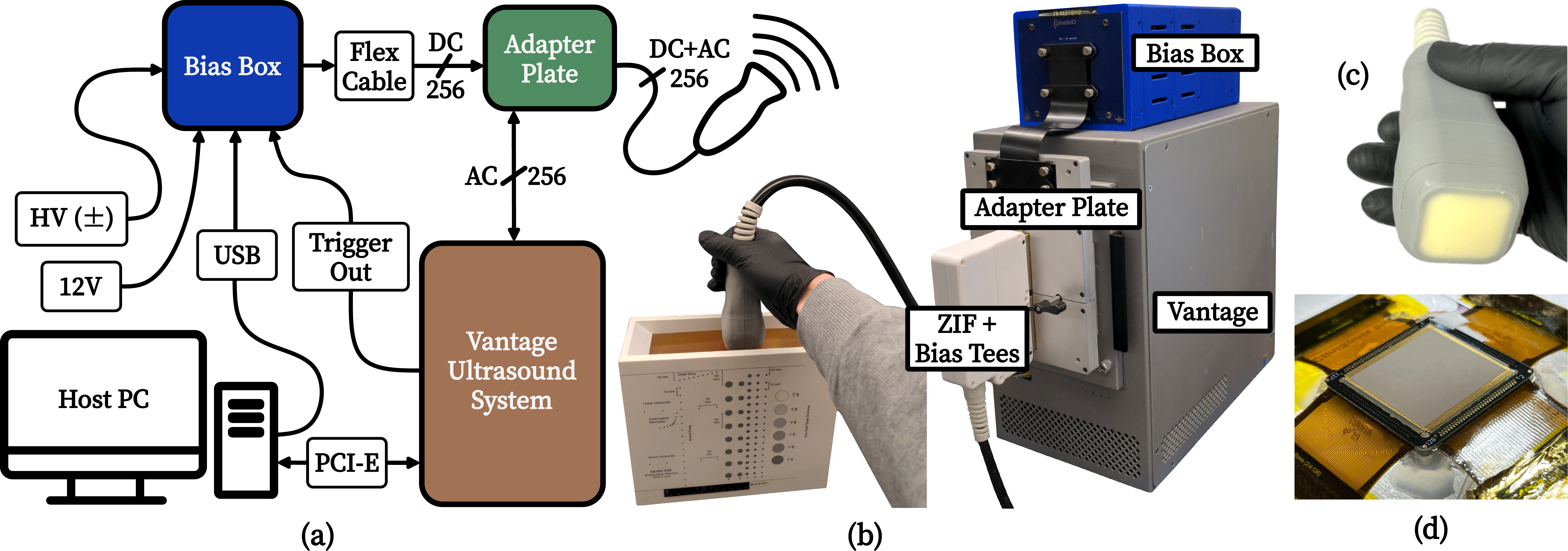}
	\caption{Handheld TOBE Array Imaging Setup. (a) System Block
	         Diagram. (b) Physical System and Imaging Phantom. (c)
	         19mm$\times$19mm 7.8MHz imaging probe. (d) Unpackaged
	         transducer. The transducer is mounted to a rigid flex
	         PCB with 2 connectors routing rows and 2 connectors
	         routing columns.}
	\label{fig:system_diagram}
\end{figure*}

\IEEEPARstart{T}{he} field of 3D medical ultrasound has seen rapid
advancements in recent years. However, current 3D imaging
techniques are hindered by channel count limitations and scanning
methods. State-of-the-art fully-wired 2D matrix arrays are
typically limited to 32$\times$32 addressable elements, a
constraint imposed by the technical complexities involved in fully
wiring larger array sizes. Larger 2D-arrays make use of
micro-beamformers, and rely on beamforming approximations, but
cannot be easily scaled. These limitations pose a significant
challenge in achieving the desired image quality, large
field-of-view, and resolution required for advanced 3D medical
ultrasound applications. Row-column arrays (RCAs) simplify wiring,
which enables the manufacturing of arrays with large element
counts, but are unable to image outside the shadow of the aperture
and are limited to one-way focusing. Bias-Sensitive RCAs, also
known as Top Orthogonal to Bottom Electrode (TOBE) arrays, enable
new imaging schemes which use bias voltage encoding to effectively
sub-divide long elements in the RCA and achieve transmit and
receive focusing everywhere with a steerable focal plane
\cite{cc17_fast, MRS22_uFORCES}. Unlike conventional row-column
arrays, bias-sensitive arrays require additional electronics.
These electronics are used to bias the TOBE arrays to turn
elements on or off or to electronically program the effective
polarity of these elements. This is possible because TOBE arrays
are fabricated from electrostrictive relaxors which are not
inherently piezoelectric until a bias voltage is applied. These
relaxors (such as PMN-PT) exhibit giant dielectric constants as
large as 20,000 which affords a strong electrostriction effect
\cite{park97relaxor,zhang10pmnpt}. The polarity of the applied
bias field induces a polarization in the relaxors aligned with the
applied field which in turn effects the polarization of
transmitted or received signals \cite{latham18simu}. In this work
we present handheld TOBE arrays with associated biasing
electronics and adapt them to a programmable research ultrasound
platform. We investigate the performance of these arrays in
comparison to conventional row-column array imaging methods
including Tilted Plane Wave (TPW) compounding and Virtual Line
Source (VLS) imaging. Using Fast Orthogonal Row Column Electronic
Scanning (FORCES) \cite{cc17_fast} we show significant image
quality improvements and imaging beyond the shadow of the aperture
without utilizing the recently demonstrated diverging lens
strategy \cite{salari25}. In contrast to that work, where the
authors demonstrated that a resolution degradation occurred due to
the use of the lens, the method we present here is free from such
artifacts and achieves transmit and receive focusing everywhere in
an electronically definable scan plane. Results are demonstrated
both in phantoms and in an animal model.

\section{METHODS}

\begin{figure*}
	\centering
	\includegraphics[width=0.9\textwidth]{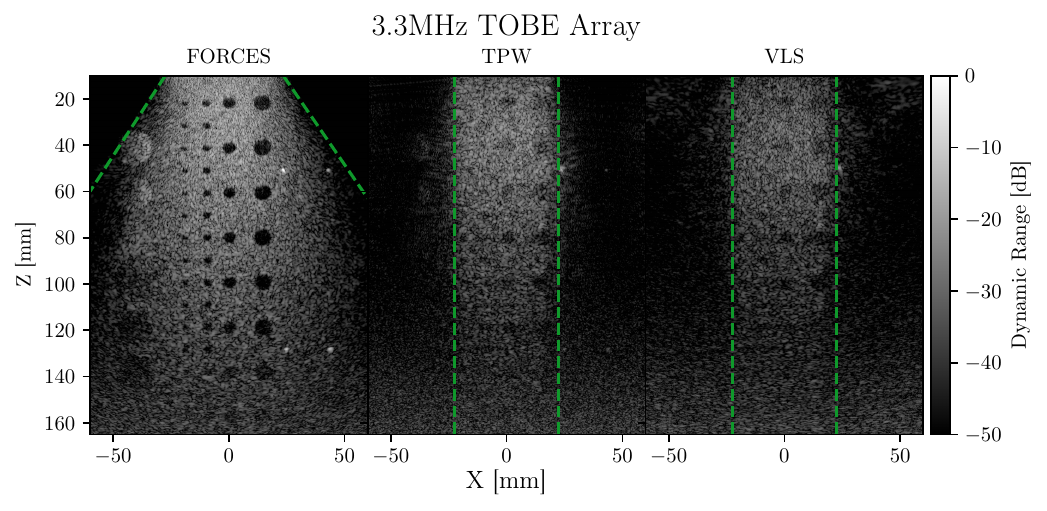}
	\\
	\includegraphics[width=0.9\textwidth]{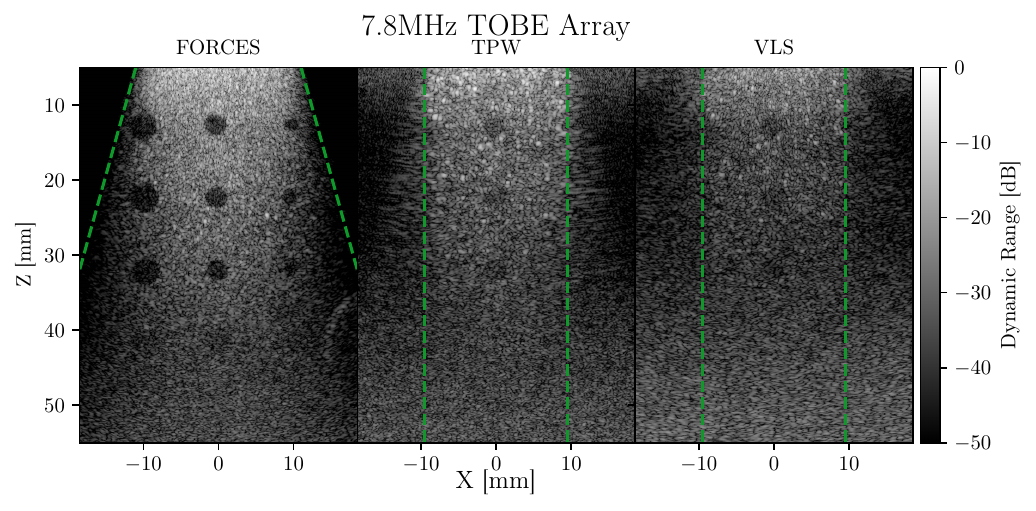}
	\caption{Imaging phantom anechoic cyst targets. The lower frequency 3.3MHz TOBE
	         is able to resolve targets down to 14cm of depth using the FORCES method
	         and is able to clearly resolve features far outside the shadow of the aperture.
	         When the 7.8MHz TOBE array was used with the FORCES method we are able to clearly
	         resolve 4, 3, and 2mm cysts. The TPW and VLS methods are only able to resolve
	         the 3mm cysts located in the center of the arrays shadow.}
	\label{fig:cysts}
\end{figure*}

\subsection{TOBE Arrays and Electronics}

We evaluated CliniSonix (Edmonton, Canada) 128$\times$128 handheld
lambda-pitch TOBE arrays with frequencies of 3.3MHz and 7.8MHz.
These arrays were adapted to a Verasonics (Kirkland, USA) Vantage
System using a CliniSonix adapter plate. The adapter plate accepts
256 bias voltage channels from the CliniSonix bias-voltage unit.
It includes two 260-pin Cannon ZIF connectors which connect
directly to the Universal Transducer Adapter on the Vantage
system. The plate routes all biasing channels and Vantage
transmit-receive channels to a pair of QLC Cannon ZIF connectors
which accept the custom ZIF connector block attached to the
probes.  This ZIF connector contains bias tees which combine the
DC biases with the transmit-receive channels. To conduct imaging,
a sequence of bias patterns is uploaded from the Vantage host PC
to the CliniSonix bias box over a USB connection. During the
imaging sequence, the Vantage generates a trigger out signal,
which is used to switch between bias patterns. After sending the
trigger, the Vantage waits for a period of time to allow the new
bias voltage to stabilize, and then it performs a transmit and
image acquisition.  Figure \ref{fig:system_diagram}(a) shows a
block diagram demonstrating all connections involved in the
system. Figure \ref{fig:system_diagram}(b) shows a photograph of a
handheld TOBE array, adapter plate, biasing box, Vantage system,
and imaging phantom. Figures \ref{fig:system_diagram}(c) and
\ref{fig:system_diagram}(d) show a fully assembled transducer and
a dissabmled tranducer without house respectively.

\begin{figure}
	\centering
	\includegraphics[width=\columnwidth]{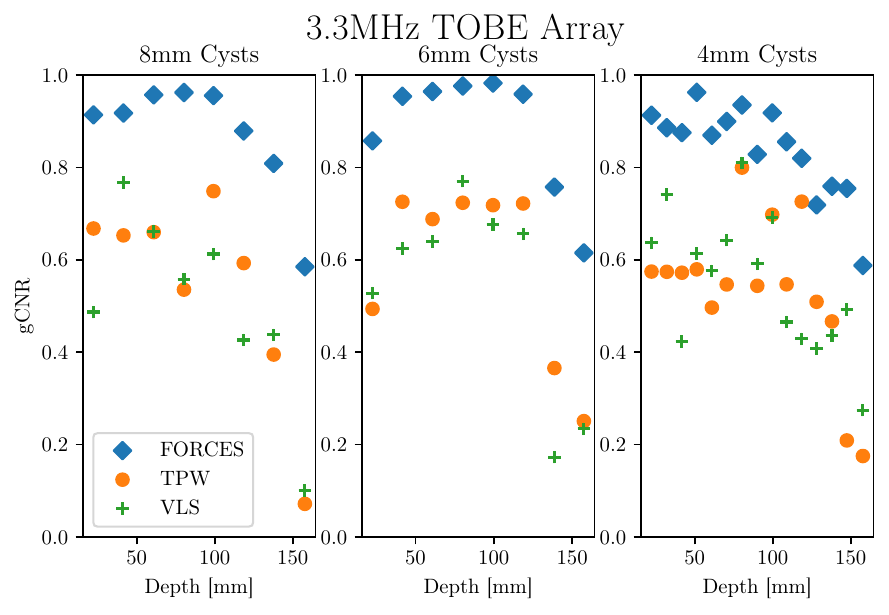}
	\\
	\includegraphics[width=\columnwidth]{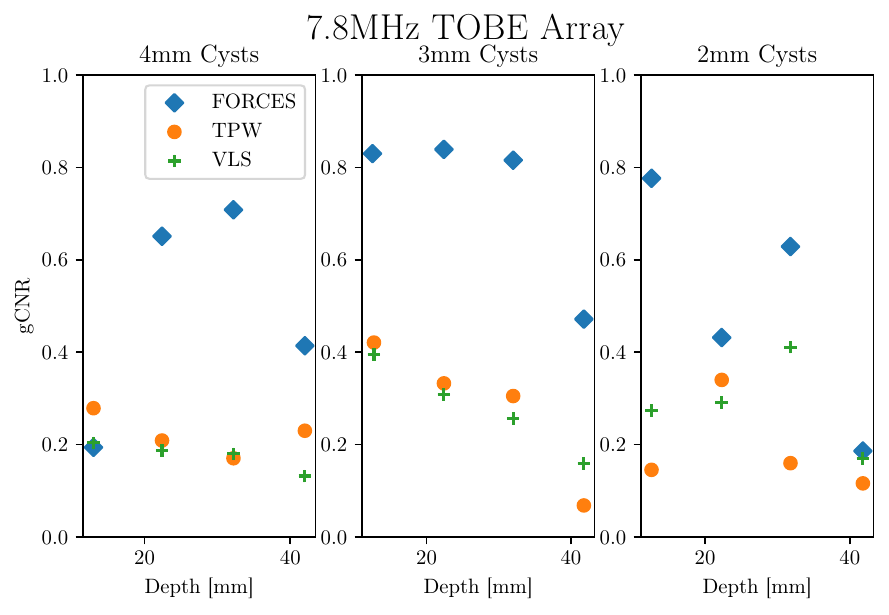}
	\caption{Cyst Generalized Contrast to Noise ratio comparison. Cysts for
	         each array were taken from Figure \ref{fig:cysts}.}
	\label{fig:gcnr}
\end{figure}


\begin{figure*}
	\centering
	\includegraphics[width=0.9\textwidth]{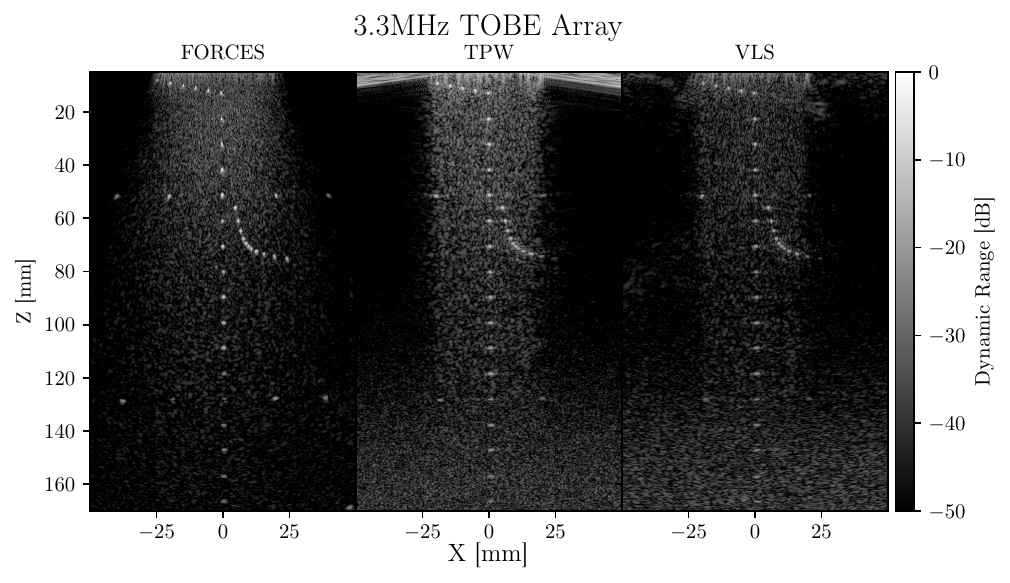}
	\\
	\includegraphics[width=0.9\textwidth]{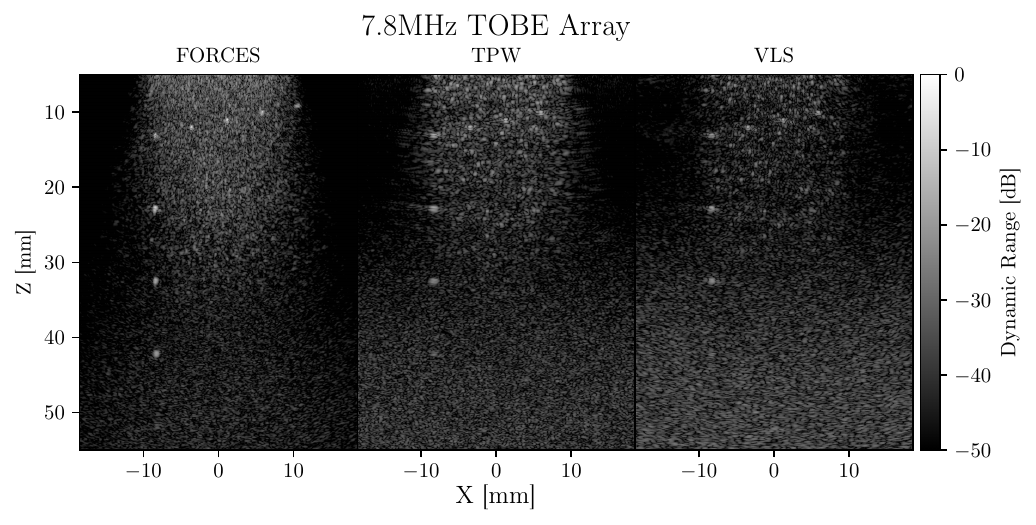}
	\caption{Phantom 100$\mu$m wire targets. The FORCES method consitently achieves
	         finer lateral resolution than VLS and TPW as well as exposing wires
	         beyond those visible with the conventional methods. Particularly
	         of note with the 7.8MHz array is FORCES' ability to clearly resolve wires
	         very close to array whereas VLS and TPW tend to lose those near wires in the
	         background speckle.}
	\label{fig:res}
\end{figure*}

\begin{figure}
	\centering
	\includegraphics[width=\columnwidth]{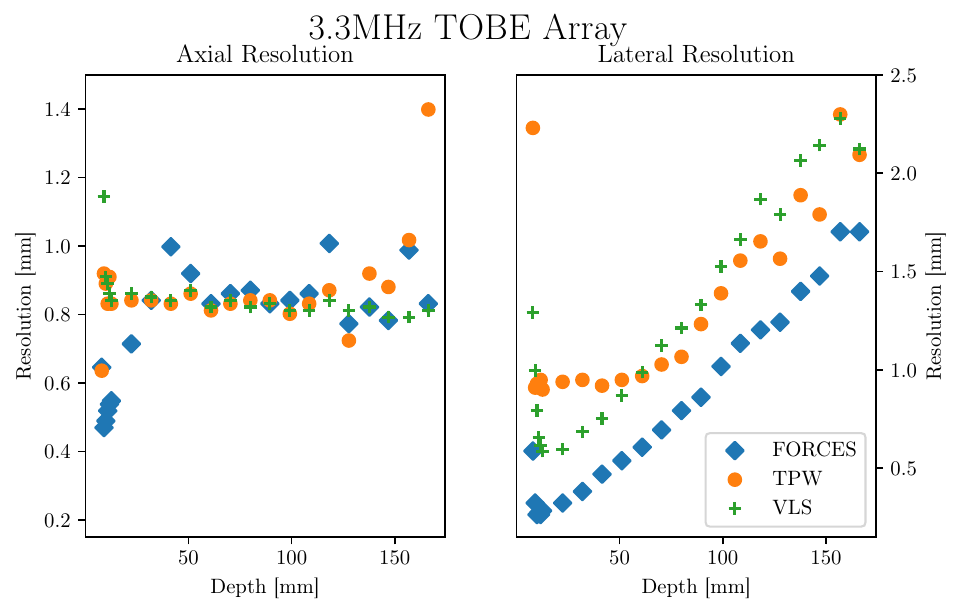}
	\\
	\includegraphics[width=\columnwidth]{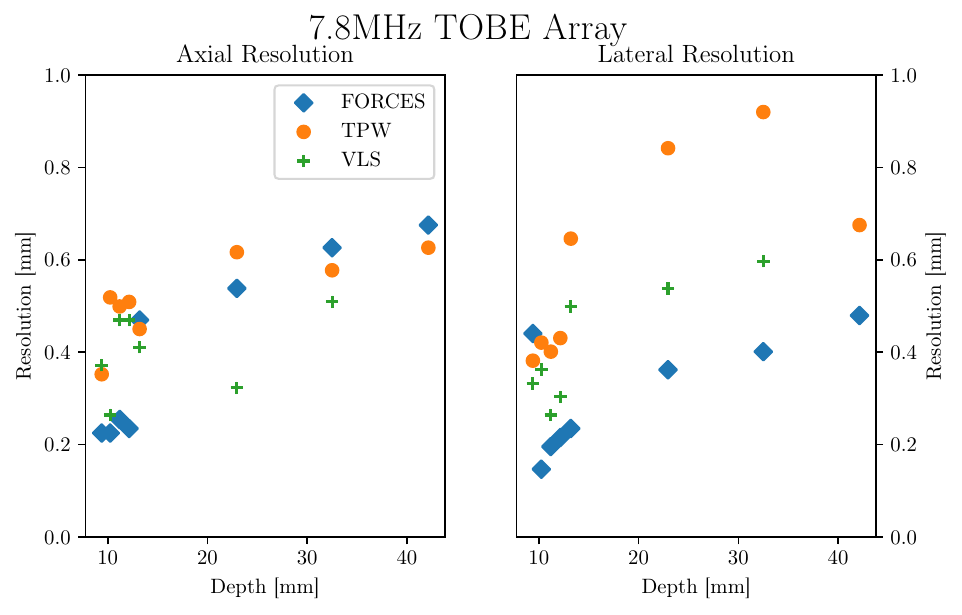}
	\caption{Comparison of measured resolutions for each imaging method. The axial
	         resolution is comparable for all methods since the same array was utilized
	         for all of them. The FORCES method consistently provides a superior
	         lateral resolution than the other methods.}
	\label{fig:res_comp}
\end{figure}

\subsection{Imaging Methods}

The study utilized the Vantage system configured to execute the
FORCES imaging sequence, as detailed in \cite{cc17_fast}. In
brief, the FORCES technique emits an elevationally focused
transmit signal along the rows of the transducer array while
applying column-wise biasing voltages derived from the columns of
a Hadamard matrix. This spatial encoding technique
\cite{chiao1997sparse} allows for the full use of the entire
aperture for each transmit. The magnitude of the bias voltages is
used to tune the sensitivity of the array, as described in
\cite{cc17_fast}. The FORCES approach can be considered an
aperture-encoded version of a synthetic aperture imaging sequence.
The aperture-decoded data is equivalent to transmitting with a
single elevationally focused column and receiving with all
columns. Therefore received signals are recorded from the columns
of the array. After inverting the polarity of negatively biased
signals and a software decoding step, an image is formed by
synthetic transmit-receive aperture processing. For a
128$\times$128 TOBE array, like the arrays used in this study, 128
transmissions are required to form a B-scan image. A recent
Ultrafast FORCES (uFORCES) \cite{MRS22_uFORCES} approach is
capable of imaging with far fewer transmit events but we do not
investigate that in this paper. The B-scan image plane is
controlled electronically using elevational focusing and beam
steering.

To perform conventional row-column array imaging, we first need to
convert our TOBE arrays into traditional row-column arrays. This
can be achieved by applying the same bias voltage to all elements.
We then implemented two different methods: the Tilted Plane Wave
(TPW) method \cite{flesch17} and the traditional multi-element
Virtual Line Source (VLS) method \cite{vls2002, rasmussen15three}.
The TPW method involves transmitting multiple plane waves with
varying emission angles on the rows while receiving on all columns
(or vice versa). This allows for dynamic receive focusing in the
azimuth direction and synthetic transmit focusing in the elevation
direction. The VLS method involves transmitting diverging waves
from virtual line elements located behind the aperture. These line
sources are walked by row transmit delays while receiving on all
columns (or vice versa). This allows for monostatic synthetic
aperture focusing in the elevation direction and dynamic receive
focusing in the azimuth direction.

We utilized a commercial quality assurance phantom, ATS-539 (Sun
Nuclear, FL, USA), for resolution and cyst targets. Acquired
images were analyzed for resolution and cyst Generalized
Contrast-to-Noise Ratio (gCNR) \cite{rodriguez20}.

\section{RESULTS}

\begin{figure}
	\centering
	\includegraphics[width=0.35\columnwidth]{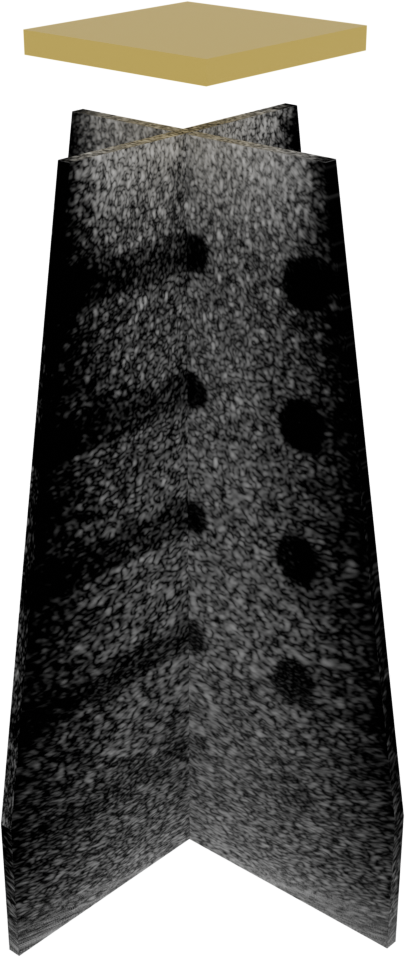}
	\caption{Cross Plane FORCES image. FORCES can be
	electronically scanned and focused to produce high
	resolution volumetric data.}
	\label{fig:xplane}
\end{figure}

\subsection{Cyst Contrast Imaging}

Figure \ref{fig:cysts} presents a comparison between the cyst
contrast achieved by the FORCES imaging method and that achieved
by the conventional VLS and TPW techniques. To ensure a fair
comparison, we compounded 128 transmits for both VLS and TPW, as
this was the necessary number of transmits to generate the FORCES
image. As shown in the figures, the FORCES method is able to
extend the field of view beyond the shadow of the aperture, which
is a major limitation of the conventional techniques.
Additionally, the FORCES images exhibit higher overall contrast,
quantified using the generalized contrast-to-noise ratio (gCNR).
gCNR values were calculated for multiple cysts across the image,
with results compiled into the graphs presented in Figure
\ref{fig:gcnr}. An example of the utilized regions of interest for
contrast calculations is provided in Supplementary Figure 1. For
the low-frequency array, the sufficient penetration depth enabled
VLS and TPW to resolve cysts at depths up to 12cm, albeit with
significantly lower contrast compared to the FORCES method.
However, when the higher frequency array was used the unfocused
transmissions from VLS and TPW failed to generate sufficient
energy to adequately resolve cysts at the edge of the array’s
field of view. As shown in Figure \ref{fig:cysts}, these
techniques exhibited limited usability at higher transmit
frequencies. For example, with the 7.8MHz TOBE array, the FORCES
gCNR for 3mm cyst targets was greater than 0.8 compared to values
typically less than 0.4 for the VLS and TPW methods for depths up
to 30mm. Similar trends were seen for the lower frequency array
and for other cyst targets.

\begin{figure*}
	\centering
	\includegraphics[width=0.85\textwidth]{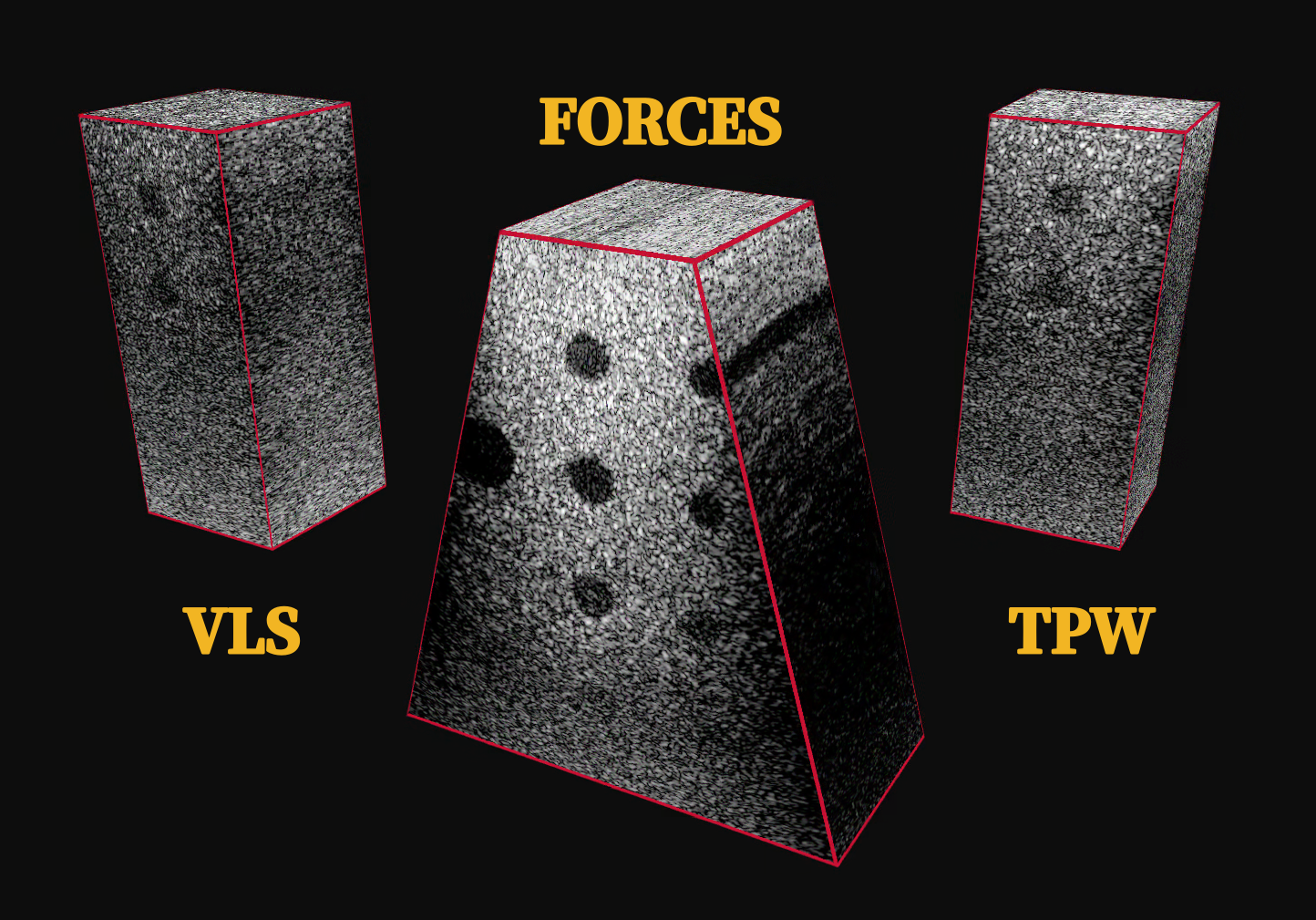}
	\caption{Experimentally acquired volumetric phantom data
	         using FORCES (center), compared with VLS and TPW (left and right).
	         Here the 7.8MHz TOBE array was used. The FORCES acquisition is not
	         possible with conventional RCAs but is possible with our TOBE arrays
	         and electronics. Note the imaging beyond the shadow of the aperture
			 and improved contrast with our approach. A video of the volumes rotating
			 is included in the supplementary materials.}
	\label{fig:volumes}
\end{figure*}

\subsection{Resolution}

The imaging phantom's 100$\mu$m wire targets were utilized to
characterize the spatial resolution achievable with each evaluated
imaging method. We report resolution as the
Full-Width-at-Half-Maximum (FWHM) linewidth of the wire targets
after beamforming. The regions of interest captured within the
imaging phantom are displayed in Figure \ref{fig:res}. For the low
frequency array, we selected the diagonal wires at the top of the
phantom as well as the wires running down the center of image. For
the high frequency array, we utilized all visible wire targets
within the field of view. The results are summarized in Figure
\ref{fig:res_comp}. As the axial resolution of these imaging
methods is primarily determined by the array's bandwidth
\cite{RSCfbmus}, the performance across the different methods was
found to be largely comparable. However in lateral direction, the
FORCES imaging technique demonstrated a consistent $\approx$25\%
improvement in lateral resolution compared to the other methods.
This advantage can be attributed to the FORCES method's ability to
leverage two-way receive focusing, whereby the transmitted beam is
dynamically focused during both the transmit and receive phases of
the imaging process. This two-way focusing approach allows for
tighter lateral beam formation and, consequently, enhanced lateral
resolution and image quality. This improvement could be valuable
for applications requiring high-precision spatial mapping, such as
small animal imaging or intravascular ultrasound visualization.

\subsection{Volumetric Imaging}

\begin{figure}
	\centering
	\includegraphics[width=\columnwidth]{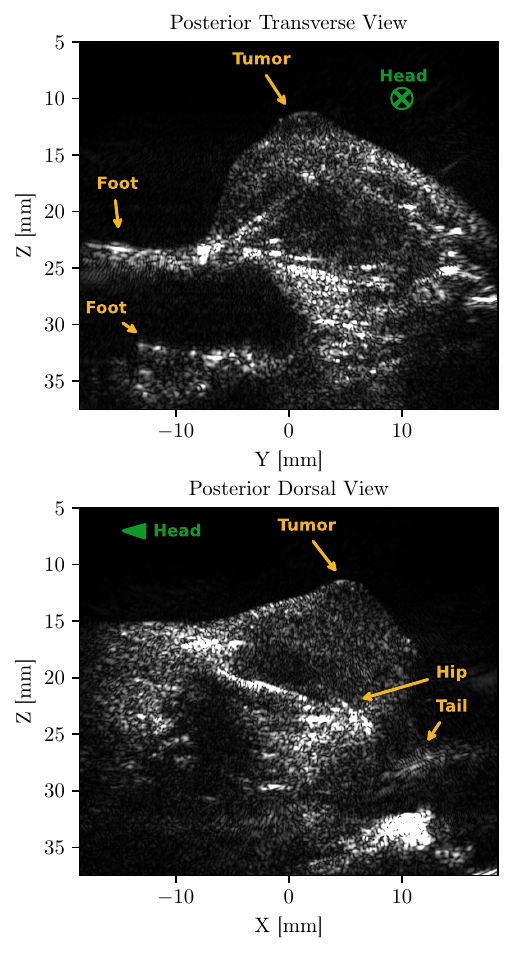}
	\caption{Mouse model with tumor. Here we show a cross-plane of the
	         rear flank acquired using an 8.3MHz TOBE Array. The mouse was
	         partially submerged in a water tank with the array located at the
	         water's surface.}
	\label{fig:mouse}
\end{figure}

The VLS and TPW methods are both volumetric imaging techniques,
whereas the FORCES method was originally designed for B-scan
imaging. Notably, as indicated by the Electronic Scanning
component of its name, the focal plane can be adjusted
electronically. This adjustment can be achieved through two modes.
The first mode involves swapping the scan plane by interchanging
the roles of the rows and columns, enabling the production of
Cross Plane images, as illustrated in Figure \ref{fig:xplane}.
This process necessitates only one additional FORCES acquisition,
making it feasible for implementation in live navigation
scenarios, although this application is not demonstrated in this
study. In clinical practice, real-time interpretation of 3D data
by sonographers is challenging; thus, high-quality cross plane
FORCES images would be advantageous for this purpose. Once a
region of interest is identified, one of the volumetric methods
can be employed for offline analysis as is common practice
\cite{tonni2015role}. Additionally, the scanning plane can be
traversed in a manner similar to walking aperture techniques. A
distinguishing feature of FORCES is that the entire aperture is
utilized for each transmission, resulting in images with enhanced
contrast, as demonstrated in previous figures. Individual planes
can subsequently be stitched together using software to create a
volumetric representation. An example of such a volume is
presented in Figure \ref{fig:volumes} with a video provided in the
supplementary material. Here we utilized 64 individual FORCES
images spatially separated by 300$\mu$m. Transmit delays were
adjusted to maintain a constant focal depth of 35mm. It is
important to remember that this FORCES volumetric image requires a
significant number of transmit events and that the final volume is
composed of discrete scan planes. We previously demonstrated
steering the focal plane in elevation between transmits
\cite{CC19_FORCES}, however, this produced significant grating
lobes so we didn't replicate that method here.

\subsection{Small Animal Imaging}

We conducted a further experiment by imaging a mouse model
\emph{ex-vivo} a TOBE Array operating at 8.3MHz. Animal
experiments were approved through the University of Alberta ACUC
(AUP \#3982 and \#2994). SCID Hairless Outbred (SHO, Charles
River) mice were inoculated with subcutaneous B16F10 mouse
melanoma tumors in the hind flank. Live ultrasound imaging was
performed on mice with their hind portion submerged in degassed
water. Cross-plane B-scans are shown in Figure \ref{fig:mouse}.

\begin{FlushLeft}
\begin{table*}
	\centering
	\caption{Comparison of TOBE FORCES vs Traditional Row-Column Imaging Methods}
	\label{tab:summary}
	\begin{tabular}{|p{10em}|p{10em}|p{10em}|p{10em}|p{10em}|}\hline\rowcolor{tablecolor!50}
	                                                       & TPW                                                                    & VLS                                                                    & FORCES B-Scan                                            & FORCES Volume \\ \hline\rowcolor{tablecolor!10}
	Acronym                                                & Tilted Plane Wave                                                      & Virtual Line Source                                                    & Fast Orthogonal Row Column Electronic Scanning           & Fast Orthogonal Row Column Electronic Scanning      \\ \hline\rowcolor{tablecolor!20}
	Reference                                              & \cite{flesch17}                                                        & \cite{vls2002, rasmussen15three}                                       & \cite{cc17_fast, MRS22_uFORCES}                          & This Work                                           \\ \hline\rowcolor{tablecolor!10}
	\# Tx Events for $N \times N$ array                    & $\sim N$                                                               & $\sim N$                                                               & As few as 4 for uFORCES, $\sim N$ for FORCES             & $N_{\text{FORCES}}*M$ (where $M$ is number of elevational planes) \\ \hline\rowcolor{tablecolor!20}
	\multirow{2}{2em}{Pros}                                & - Fast volume imaging                                                  & - Fast volume imaging                                                  & - Transmit and receive focusing everywhere in scan plane & - Transmit and receive focusing everywhere in scan plane \\\rowcolor{tablecolor!20}
	                                                       &                                                                        &                                                                        & - Imaging beyond shadow of aperture                      & - Imaging beyond shadow of aperture                      \\ \hline\rowcolor{tablecolor!10}
	\multirow{2}{2em}{Cons}                                & - Cannot image outside shadow of aperture                              & - Cannot image outside shadow of aperture                              & - Single elevation focus per FORCES sequence             & - Single elevation focus per FORCES sequence             \\ \rowcolor{tablecolor!10}
	                                                       & - One-way (transmit or receive) focusing in both elevation and azimuth & - One-way (transmit or receive) focusing in both elevation and azimuth &                                                          & - Long acquisition time                                  \\ \hline\rowcolor{tablecolor!20}
	Can be done with conventional RCA                      & Yes                                                                    & Yes                                                                    & No                                                       & No                                                       \\ \hline\rowcolor{tablecolor!10}
	Hardware needed (besides research ultrasound platform) & RCA or TOBE + Biasing Electronics                                      & RCA or TOBE + Biasing Electronics                                      & TOBE + Biasing Electronics                               & TOBE + Biasing Electronics                               \\ \hline\rowcolor{tablecolor!20}
	\end{tabular}
\end{table*}
\end{FlushLeft}

\section{DISCUSSION AND CONCLUSION}

The results presented here demonstrate the enhanced image quality
and expanded field of view offered by TOBE array imaging and the
associated FORCES imaging methods, in comparison to traditional
RCA imaging techniques. The added addressability provided by
biasing allows for azimuthal synthetic transmit-receive aperture
imaging after aperture decoding. This capability enables effective
transmit and receive focusing throughout the scan plane with high
signal-to-noise ratio (SNR), a significant improvement over
conventional RCA technology, which lacks the ability to subdivide
long rectangular elements. Furthermore, our approach facilitates
wider field of view imaging beyond the shadow of the aperture,
unlike conventional RCAs, where the long rectangular elements
restrict directivity to areas directly beneath the aperture's
shadow. This ability to image beyond the shadow of the aperture is
expected to be crucial for various future clinical applications.
While other groups have explored the use of RCAs to extend imaging
beyond the shadow of the aperture through diverging acoustic
lenses \cite{salari25}, their methods do not provide the synthetic
transmit-receive focusing advantages that our approach offers.
The advantages and disadvantages of the different imaging methods
are summarized in Table \ref{tab:summary}. While FORCES volume
acquisitions require more transmit events than VLS and TPW, the
improved image quality and field of view provided by FORCES can be
an attractive alternative. It is worth noting that conventional
RCA approaches, including VLS and TPW, can be implemented using
our TOBE arrays and electronics by simply applying a constant DC
voltage to all rows or columns. However, the new FORCES imaging
schemes cannot be implemented with RCAs. Recent research has
demonstrated a variety of new RCA imaging methods for applications
such as microvessel imaging \cite{song23super, lok24ultrasound},
vector flow imaging \cite{david23new, haniel23efficacy}, molecular
imaging \cite{heiles21advent, heiles25nonlinear}, functional
imaging \cite{rabut20pharmaco, bertolo21whole, morisset22retinal},
ultrasound localization microscopy \cite{errico15ultrafast,
lowerison22aging}, and shear wave elastography \cite{song12cuse,
huang20three, dong24three}. Our array technology should be able to
achieve these and other advances, given the improved image quality
it provides. Implementing the FORCES imaging schemes requires
additional biasing electronics that are not normally needed for
RCA imaging, but these are now commercially available (CliniSonix
Inc. Edmonton, AB, Canada).

Future work could explore ways to improve the elevational focusing
capabilities of the method and investigate how the expanded field
of view and higher resolution enabled by the method benefit
ultrafast imaging modalities such as 3D power doppler, vector flow
imaging, and elastography. Additionally, the recently developed
Ultrafast FORCES (uFORCES) technique \cite{MRS22_uFORCES} could be
utilized to enhance imaging speed by leveraging sparse
transmitting methods enabled by unique bias pattern groupings.
Future studies should also assess the system's sensitivity to
motion and explore potential motion compensation algorithms.
Finally, upcoming work should aim to demonstrate new TOBE
array-enabled imaging methods that can produce high-quality
volumetric data using fewer transmits than the demonstrated
walking FORCES technique.

In summary bias-switchable TOBE array technology appears to offer
major advantages compared to traditional RCA technology. The
potential to scale these arrays to unprecedented sizes while
maintaining modest channel counts provides many exciting
opportunities for next generation clinical ultrasound.

\section*{CONFLICTS OF INTEREST}

RJZ and MRS are directors and shareholders of CliniSonix Inc.,
which provided partial support for this work.  RJZ is a founder
and director of OptoBiomeDx Inc., which, however, did not support
this work. RJZ is also a founder and shareholder of IllumiSonix
Inc., which, however, did not support this work. JB is a
shareholder and director of SoundBlade Inc., and DaxSonics Inc,
which, however, did not support this work.

\section*{REFERENCES}
\printbibliography

\vfill\pagebreak

\end{document}